\documentclass[10pt, conference, letterpaper]{IEEEtran}
\IEEEoverridecommandlockouts
\usepackage{cite}
\usepackage{amsmath,amssymb,amsfonts}
\usepackage{algorithmic}
\usepackage{graphicx}
\usepackage{textcomp}
\usepackage{xcolor}
\usepackage[normalem]{ulem}
\usepackage[ruled,vlined]{algorithm2e}
\usepackage{caption}
\usepackage{subcaption}

\def\BibTeX{{\rm B\kern-.05em{\sc i\kern-.025em b}\kern-.08em
    T\kern-.1667em\lower.7ex\hbox{E}\kern-.125emX}}
    
    \renewcommand{\baselinestretch}{.94}
\begin{document}

\title{A Real-time Framework for Trust Monitoring in a Network of Unmanned Aerial Vehicles 
}


\author{
	\IEEEauthorblockN{
	Mahsa Keshavarz\IEEEauthorrefmark{1}, Alireza Shamsoshoara\IEEEauthorrefmark{1}, Fatemeh Afghah\IEEEauthorrefmark{1}, Jonathan Ashdown \IEEEauthorrefmark{2}
	}
	\IEEEauthorblockA{\IEEEauthorrefmark{1}School of Informatics, Computing and Cyber Systems, Northern Arizona University, Flagstaff, AZ, USA\\
	 Email: \{mk959,alireza\_shamsoshoara,fatemeh.afghah\}@nau.edu}

	\IEEEauthorblockA{\IEEEauthorrefmark{2} Computer Information Systems Department, SUNY Polytechnic Institute, Utica, NY, USA,\\ E-mail: ashdowj@sunypoly.edu}}
\maketitle


\begin{abstract}
Unmanned aerial vehicles (UAVs) have been increasingly utilized in various civilian and military applications such as remote sensing, border patrolling, disaster monitoring, and communication coverage extension. However, there are still prone to several cyber attacks such as GPS spoofing attacks, distributed denial-of-service (DDoS) attacks, and man-in-the-middle attacks to obtain their collected information or to enforce the UAVs to perform their requested actions which may damage the UAVs or their surrounding environment or even endanger the safety of human in the operation field. In this paper, we propose a trust monitoring mechanism in which a centralized unit (e.g. the ground station) regularly observe the behavior of the UAVs in terms of their motion path, their consumed energy, as well as the number of their completed tasks and measure a relative trust score for the UAVs to detect any abnormal behaviors in a real-time manner. Our simulation results show that the trust model can detect malicious UAVs, which can be under various cyber-security attacks such as flooding attacks, man-in-the-middle attacks, GPS spoofing attack in real-time.
\end{abstract}


\section{Introduction}
Unmanned aerial vehicles (UAVs) recently play a major role in several civilian and military operations including remote sensing, surveillance, package delivery, and medical services  \cite{rahmati2019dynamic,hodgson2016precision,eldosouky2017resilient,shamsoshoara2019solution,shamsoshoara2019distributed,shamsoshoara2019autonomous}.  Their unique features including high-mobility, ease of deployment, and their ability to hover enable them to provide services in hard-to-reach regions or time-critical missions during man-made and natural disasters in order to provide urgent Internet and communication services when necessary or for imaging purposes \cite{mozaffari2018beyond,Afghah_INFOCOM19}.

Despite this wide range of often critical UAV missions, the UAV networks are still vulnerable to several attacks including cyber-attacks such as false data injection \cite{challita2019machine}, physical attacks such as targeting the UAVs using firearms \cite{eldosouky2019drones}, and cyber-physical attacks such as GPS spoofing \cite{altawy2016security}. Therefore, an important step toward safe applications of UAVs networks is developing robust trust monitoring mechanisms to identify potential attacks on these systems during the flight. 

There are several existing trust management mechanisms as further discussed in Section \ref{sec:relatedworks} to detect the malicious UAVs.
However, the majority of these methods either focus on a specific type of aircraft, or an specific type of cyber-security attacks. Another challenge with several current proposed methods is that they are not able detect the attacks in real-time, since they often look at the history of the UAVs' trust to evaluate if they have deviated from normal behavior long enough to impact their reputation \cite{Afghah_ACC18}. It means that if the UAVs with a good reputation are targeted by an attacker, the attack may not be detected immediately. The reputation-based methods also involve storing the reputations of all agents over the course of time at the central unit, which require a large memory space.  Another trend in trust monitoring is the class of cooperative trust monitoring methods, where all the agents coordinate with one another to monitor the behavior of their teammates. While such methods are robust against the "single point of failure" problem, where the centralized trust monitoring unit may be attacked, they involve large energy consumption and computation load at the UAVs to observe other agents' behavior and also impose a heavy signaling load to the system. Some other trust monitoring approaches \cite{mcneely2016detection} were proposed based on observing the statistical measures to detect malicious UAV in real-time, however, these methods cannot distinguish between the malicious behavior or potential change of behavior due to harsh environmental conditions.

In this paper, we propose a trust monitoring framework to detect various attacks such as DDoS attack, GPS spoofing attack, man-in-the-middle attack, and the agent's potential selfish act in a network of UAVs in a real-time manner. In this method, the relative trust scores of all UAVs are regularly measured at a central unit by observing several factors including (i) the observation of the central unit from the environment to estimate how successful a UAV has been in completing its assigned tasks, (ii) the deviation of the UAVs from their original path, and (iii) the energy consumption of the UAVs.  The UAVs under the DDoS and GPS spoofing attacks consume more energy than normal UAV, and are more likely to deviate from the predicted paths. Also, the UAVs under the man-in-the-middle attack often complete a less number of assigned tasks related to a normal condition. Therefore, the selected factors to monitor the trust can capture various common attacks. The trust score of each UAV will be evaluated compared to the trust scores of its teammates to identify if a UAV has been attacked.
This method can be applied to any types of aircraft, can detect the attacks in real-time instead of relying on the history of the agents' behavior and does not require large memory or computation capabilities. Furthermore, the proposed method can differentiate the malicious behavior from the unstable behavior due to harsh environmental conditions.  


\section{Related Works} \label{sec:relatedworks}
There are several methods to monitor the safety of different aircraft such as the fault detection methods which can detect any abnormalities in the aircraft. Qi et al. \cite{qi2013literature,qi2014review} provides a comprehensive review of recent methods for fault diagnosis.
Birnbaum et al. \cite{birnbaum2016unmanned} propose a method in which the Recursive Least Squares (RLS) algorithm is applied to perform estimation and tracking of the controller parameters. In this method, data in batches of size 500 samples are processed and compared to detect discrepancies indicating anomalies. 

Several anomaly detection methods are tailored to specific types of attacks. \cite{birnbaum2016unmanned} overviews of the most popular six types of threats for UAVs, such as hardware failure, malicious hardware, and control computer attacks. In order to detect some types of hijacking or hardware failures, a Hardware Health Monitor is utilized, which monitors all of the UAV sensor data. This data consists of the positioning data and flight surfaces monitored by using the RLS algorithm. Another common attack in UAV networks is the GPS spoofing attack. In order to detect this type of attack, \cite{warner2003gps} discussed a technique, in which the receiver observes the received signal strength and compares it to the predicted signal strength over time. \cite{o2012real} proposed a method that uses two GPS receivers and check their cross-correlations to detect GPS spoofing attacks. However, this technique cannot detect the spoofing attack when the signals are weak. In order to detect the GPS spoofing attacks and determine the real locations of UAVs, \cite{eldosouky2019drones} proposes a framework, where the UAVs are allowed to travel on the shortest path between any given two locations by using the optimal UAV’s controller. This model can capture the possible GPS spoofer on the UAV’s traveling path. This method enables the UAVs to determine their real location under GPS spoofing attack by using the neighbor UAVs’ real location and their relevant distances. 

\cite{mcneely2016detection} proposed a model-based trust monitoring approach where several basic statistical measures are used to track the characteristics of a flight. These statistical measures including the mean, variance, and covariance across multiple flights at different stages of the flight are stored as a “fingerprint” of the flight, where the variations of the flight pattern can alarm that the UAV was hijacked. A threshold is set to determine at what point the flight is no longer normal. In this method, the normal flight statistics are added to the baseline and any abnormal ones are removed from the baseline. This method is tested by using 20 potential hijacking simulations and 50 safe flights. All of the obvious hijacking scenarios were detected correctly, however some unstable flights that were close to the baseline windy flights remained undetectable. 


\section{System Model}

Here, we propose a centralized trust monitoring mechanism to monitor the behavior of multiple UAVs during the flight and detect any potential abnormal behaviors in real-time. Let us consider a homogeneous network of $N$ UAVs which fly over a $M \times{ M}$ area, where $M$ is a positive arbitrary integer. $M$ is chosen randomly from $[1.5km,~2.5km]$ range. 
Each UAV is assigned with a particular task which determines the UAV's flight pattern during its operation. Some examples of these tasks include package delivery, aerial photography, and geophysical survey. It is assumed that the UAVs  are randomly distributed in the area of the network prior to the mission. It is also assumed that among the UAVs which operate in a close proximity of each other, only one UAV is under an attack in a given time \cite{eldosouky2019drones}. In proposed trust monitoring model, we calculate the trust score of each UAV and compare it with the trust scores of its neighbor UAVs. Since all the neighbor UAVs experience similar environmental conditions, if a UAV presents an out-of-range trust score it can be identified as an under-attack UAV.
This trust monitoring method can be scaled up to a large network by dividing the UAVs to several clusters, where the UAVs withing each cluster experience similar environmental conditions.


 After clustering the neighbor UAVs, the central unit (e.g., an audit unit or the ground station) regularly measure the trust score of all the UAVs within a cluster to keep track of their behavior in order to detect any abnormalities in a real-time manner. After each mission, the trust score of the UAVs is reset to zero. To calculate this \emph{trust score}, we consider three factors: the task success rate, the energy consumption rate, and the deviation of the UAV's flight trajectory from its expected path. The reason behind choosing these three factors is that most common attacks impact one or more of these behavioral factors. For instance, when a UAV is under the man-in-the-middle attack, it usually skips some assigned tasks (i.e. present a low task success rate). When a UAV is under the DDoS attack or the hijacking attack, it will present an unusual energy consumption pattern, where it may consume more energy (because the attacker forces the UAV to perform unexpected tasks) or less energy (because the attacker prevents the UAV from performing its task), or when the UAV is under the GPS spoofing which the attacker wants to mislead the UAV into another location (deviation from the predicted path). 
In continue, we define the three factors of trust.
\subsection{Task Success Rate}
One of the main factors to evaluate the trustworthiness of an agent is the task success rate. During a flight, each UAV is assigned with a couple of tasks and they are expected to complete all of these tasks. It means that in a normal condition, the number of successfully completed tasks should be higher than the number of failed tasks. Sometimes due to environmental and technical issues, the UAVs are not able to complete all of their tasks, but still, the number of failed tasks is lower than the number of successful tasks. However, when the agents are under certain attacks such as the man-in-the-middle attack, or the forwarding attack, they may deliver a less number of successful tasks.

To evaluate the behavior of UAVs in terms of performing their assigned tasks, we consider the history of completed tasks for the UAVs during each mission. The trust factor is expected to differentiate whether the underlying reason for not completing the assigned tasks is related to an attack or because the UAV is experiencing any environmental or technical issues. Therefore, the trust factor considers the relative performance of the UAVs rather than an individual performance. In this case, if the UAVs are operating under bad weather conditions, they all will underperform and deliver a low task success rate. However, if one of the UAVs is under an attack, only this UAV will experience a low task completion rate. Moreover, since the task success rate is measured upon the completion of a mission, it gives the UAVs the time to recover from the potential technical issues they may face. For example, if the UAV has a low battery level, it has the opportunity to put a hold on performing the tasks to land on a recharging spot and then complete the rest of the assigned tasks. Hence, when looking at the record of this UAV during the entire mission, the number of successful tasks is higher than the failed tasks.

The history of completed tasks can be considered either based on the direct reports of the UAVs which is prone to receiving false reports from the under-attack UAVs or it can be estimated based on the direct or indirect observation of the central unit. For instance, if the UAV's task  was to survey a specific area, the central unit can determine if the task has been successfully completed based on whether obtaining the image of the entire region or not rather than relying on the UAV's self-report. Indeed, the central unit cannot always have such a direct observation for all the assigned tasks. For instance, if the UAV's task was to drop a fire ball to ignite fire in a particular area (a mechanism used to initiate controlled fires to prevent wildfires \cite{ignite} ), then the central unit may not be able to accurately determine if the ball was dropped. However, it can estimate if the task was performed by the UAV by observing its impact on the field (i.e. fire in the target area) with some level of uncertainty.  To account for such potential uncertainty in the assessment of the central unit in terms of the number of successful or failed tasks for each UAV, we propose a trust factor based on subjective logic framework (SLF) \cite{josang1999algebra}. Let us assume that $s$ is the number of successful tasks,  $f$ is the number of unsuccessful tasks, and $x$ denotes the number of tasks that the central unit can not certainly declare as successful or unsuccessful. 
Based on this theory, the trust is composed of a vector $T = \{b,\:d,\:u \}$. The parameters $b,\:d,\:u$ respectively represent the probability of trust, the probability of distrust, and the chance of uncertainty, where they satisfy $b + d + u = 1$ and $b,\:d,\:u \in [0, 1]$.  The aforementioned parameters are calculated as follows:

\begin{align}\label{eq:trust_fraction}
b = \frac{s}{s + f + x},\:d = \frac{f}{s + f + x},\:u = \frac{x}{s + f + x}
\end{align}


The calculation of the task success rate for binary statements, which in our case, it is reliable and unreliable UAVs is as below \cite{josang1999algebra}:
\begin{align}\label{eq:trust_count}
& T_{task} =  \frac{2b + u}{2} 
\\
\nonumber
\end{align}

\subsection{Energy Consumption}
Another important factor to asses the level of trust for an agent is to calculate its energy consumption noting the expected amount of energy for its assigned task. In a normal condition, the energy consumption rate for each task should be within a certain range. However, when the UAVs are under an attack (e.g., the flooding and GPS spoofing), the energy consumption rate would be higher or lower than the normal range. Further, in bad weather conditions (e.g., strong winds), the UAVs need to consume more energy to complete their assigned tasks. In order to distinguish between the malicious conditions and environmental conditions, we compare the energy consumption of the UAVs in one cluster. For example, if the energy consumption of one UAV is higher than the other neighbor UAVs, it means that the UAV is under a flooding or a GPS spoofing attack. Also, a low or high energy consumption of a single UAV can be an indicator that the control of the UAV has been taken over by an adversary. However, under unstable environmental condition like heavy wind, all the neighbor UAVs in a cluster present an unusual  energy consumption as they all experience similar a weather condition.
Noting the assumption that the attackers can target one UAV at a time in one cluster \cite{eldosouky2019drones}, this proposed approach can reduce the rate of false alarms in reporting the UAVs as malicious in extreme environments.  

As we mentioned before, each UAV in the same cluster has the same initial energy, and since all UAVs are assigned with similar types of tasks, they are expected to have a similar energy consumption.
In order to evaluate the behavior of the UAVs within a cluster in terms of their energy consumption, we need to compare the energy consumption of each UAV (let us say UAV $i$) with the average consumption  of all other UAVs in that cluster based on (\ref{eq:energy_ave}). Let $K$ denote the number of UAVs in one cluster, and  $E_{i}$ denote the consumption  energy of the UAV $i$.
The energy trust factor of UAV $i$, $ T_{ene_i}$ can be calculated based on the absolute value of the difference between $E_{i}$ and $E_{{ave}_i}$ as shown in (\ref{eq:energy_trust}).

\begin{align}\label{eq:energy_ave}
& E_{avg_{-i}} = \frac{1}{K - 1} \sum_{j=1, j \neq i}^{K}E_{j},
\end{align}
\begin{align}\label{eq:energy_trust}
& T_{ene_i} = \frac{|E_i - E_{avg_{-i}}|}{E_{avg_{-i}}} 
\end{align}

\subsection{Path Deviation}
Each UAV in a cluster is assigned with several similar tasks during their mission, and a particular path is expected for each task before the mission is started. We assume that the UAVs should follow the expected path; however, due to the different reasons, the actual trajectory of the UAVs may be different from their expected paths. One reason can be an extreme weather condition (e.g.,  strong wind) to cause deviation from the predicted path. Also, an obstacle in the UAV's expected path can cause a deviation from the original path for a short time. Another  reason for such deviation from the expected path can be different cyber-physical attacks such as GPS spoofing. 

In order to find out the main reason behind the UAV's path deviation, the audit unit observes the current location of the UAVs based on the GPS information, and also it has an estimation of the existing obstacles in the predefined region. In order to calculate the deviation trust, the audit unit looks at the history of the actual locations for $\alpha$ consecutive time slots. At each time slot, the audit unit calculates the difference between the actual and expected location; then, it calculates the average of the differences for the $\alpha$ time slots. We can define $\alpha$ based on the size of predefined obstacles. (\ref{eq:trust_dev}) expresses the trust deviation at each time slot: 

\begin{align}\label{eq:trust_dev}
& T_{dev} = 
\\
\nonumber
& \frac{\sum\limits_{i=t-\alpha}^{t}\sqrt{(x_{ex_i} - x_{ac_i})^2 + (y_{ex_i} - y_{ac_i})^2 + (z_{ex_i} - z_{ac_i})^2}}{\alpha} ,
\end{align}
where $x_{ex},y_{ex}, z_{ex} $ is the expected location of the UAV and  $x_{ac}, y_{ac}, z_{ac}$ is the actual location at time $i$.

If the deviation from the expected path is related to facing an object or temporal harsh weather conditions, the UAV is expected to return to the original path withing a short portion of time; however, if the UAV has been attacked this deviation will be observed for a long period of time. In this case, the value of deviation trust is increased, which impacts the total trust score. Since, the trust scores are compared with one another, if the UAVs face extreme weather condition, they will all have a considerable long-term deviation from their original paths and this condition can be differentiated from potential attacks.




Environmental issues may consume more energy than normal condition and cause deviation from the expected path. As we discussed before, we compare the energy consumption of the UAV with the neighbor UAVs, which they are in the same cluster. In this case, by considering the amount of both energy consumption and the deviation, we can recognize which one is attack deviation and which one is a weather deviation.

\subsection{Calculation of Trust score}
The overall trust score of a UAV is defined by integrating the task success rate trust, the energy trust, and the deviation trust as follows: 
\begin{align}\label{eq:trust}
T_{total} = w_{task} \times T_{task} + w_{ene} \times T_{ene} - w_{dev} \times T_{dev},
\end{align}
where $w_{task}$, $w_{ene}$, and $w_{dev}$ are weights, where $w_{task} + w_{ene} + w_{dev} = 1$, $w_{task},w_{ene},w_{dev} \in [0,1]$.
We first calculate the task success rate trust, energy trust and deviation trust, then obtain the overall trust score of an agent using  (\ref{eq:trust}). Since the deviation from the actual path decreases the total trust score, we subtract the deviation from the total trust score. Then, the trust score of each UAV in one cluster is compared with the trust score of its neighbor UAVs to see whether the trust scores are in the same range or not. For example, we have three UAVs as it is shown in Algorithm~\ref{algo:algorithm_trust}. In the normal condition, the trust score of three UAVs should be in the same range. If the range of trust score of $UAV_1$ is different from the other UAVs, it means the $UAV_1$ is under an attack.


\section{Simulation Results and Analysis}
In this section, we evaluate the performance of the proposed trust monitoring model in detecting the malicious UAVs using extensive simulations. We consider different Monte-Carlo scenarios, where each scenario runs 1000 times for different types of tasks, and the average is calculated as a result. In each scenario, we assume we have three UAVs in one cluster, which means they have the same initial energy, similar type of task, and all of them are in the same environmental condition. Only one UAV in a cluster can be under an attack. In each scenario, we study one attack, which  impacts  at least one factor of the trust model. 

In this model, we monitor the trust score of the UAVs by calculating the task success rate trust, the energy trust, and the deviation trust at the ground station in a periodic way (e.g. we consider four-minute intervals). After calculating these three factors, we calculate the total trust score for each UAV and compare them to see whether they are in the same range or not. Based on the assumption that in each flight, just one of the UAVs can be under the attack, if the range of one UAV’s trust score is different from the others, that UAV is reported malicious.

\begin{figure}[hbt]
     \centering
     \begin{subfigure}[b]{\columnwidth}
         \centering
         \includegraphics[width=0.80\columnwidth]{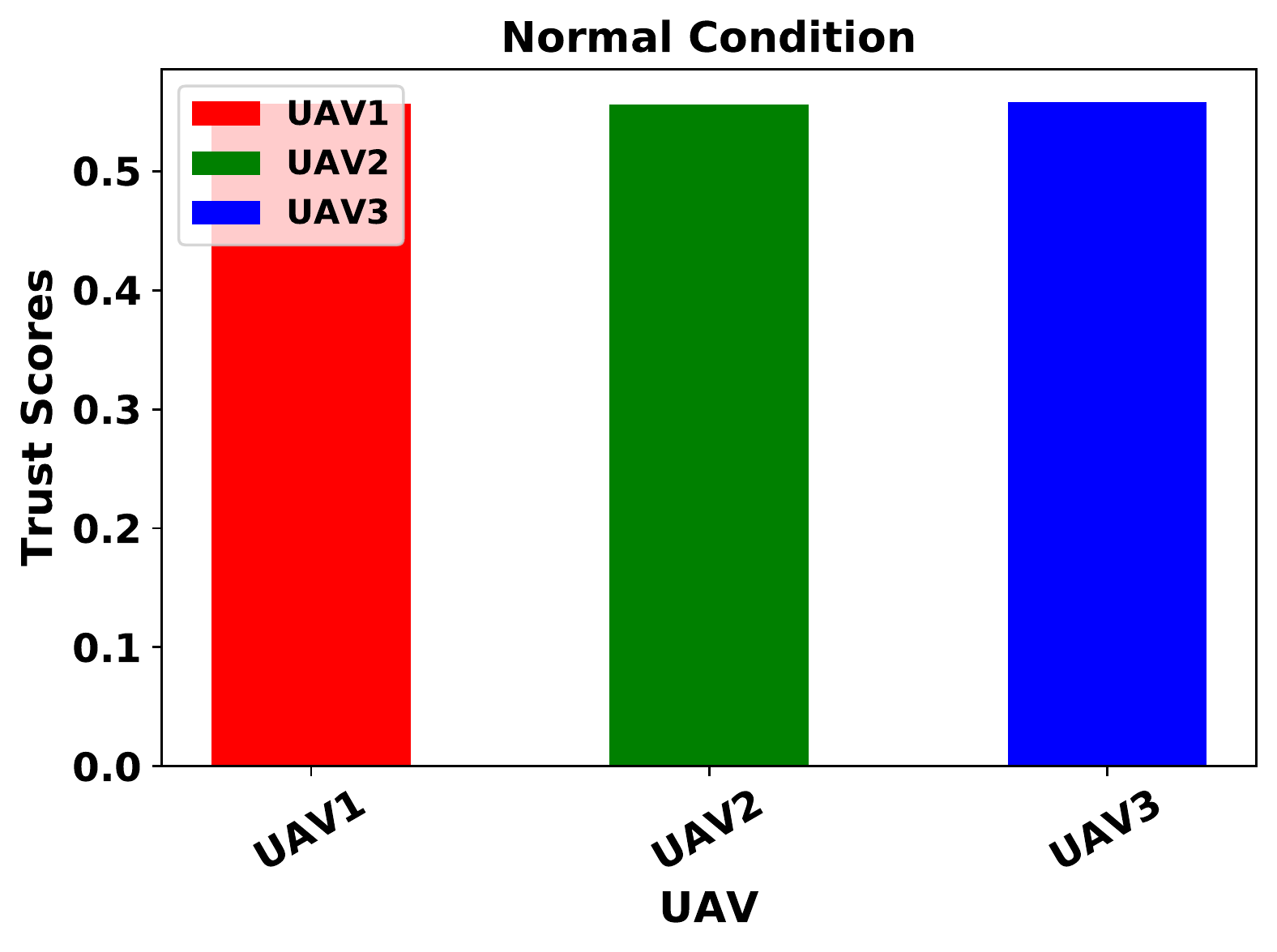}
         \caption{Trust scores of all UAVs under a normal condition.}
         \label{fig:fig2a}
     \end{subfigure}
     \begin{subfigure}[b]{\columnwidth}
         \centering
         \includegraphics[width=0.80\columnwidth]{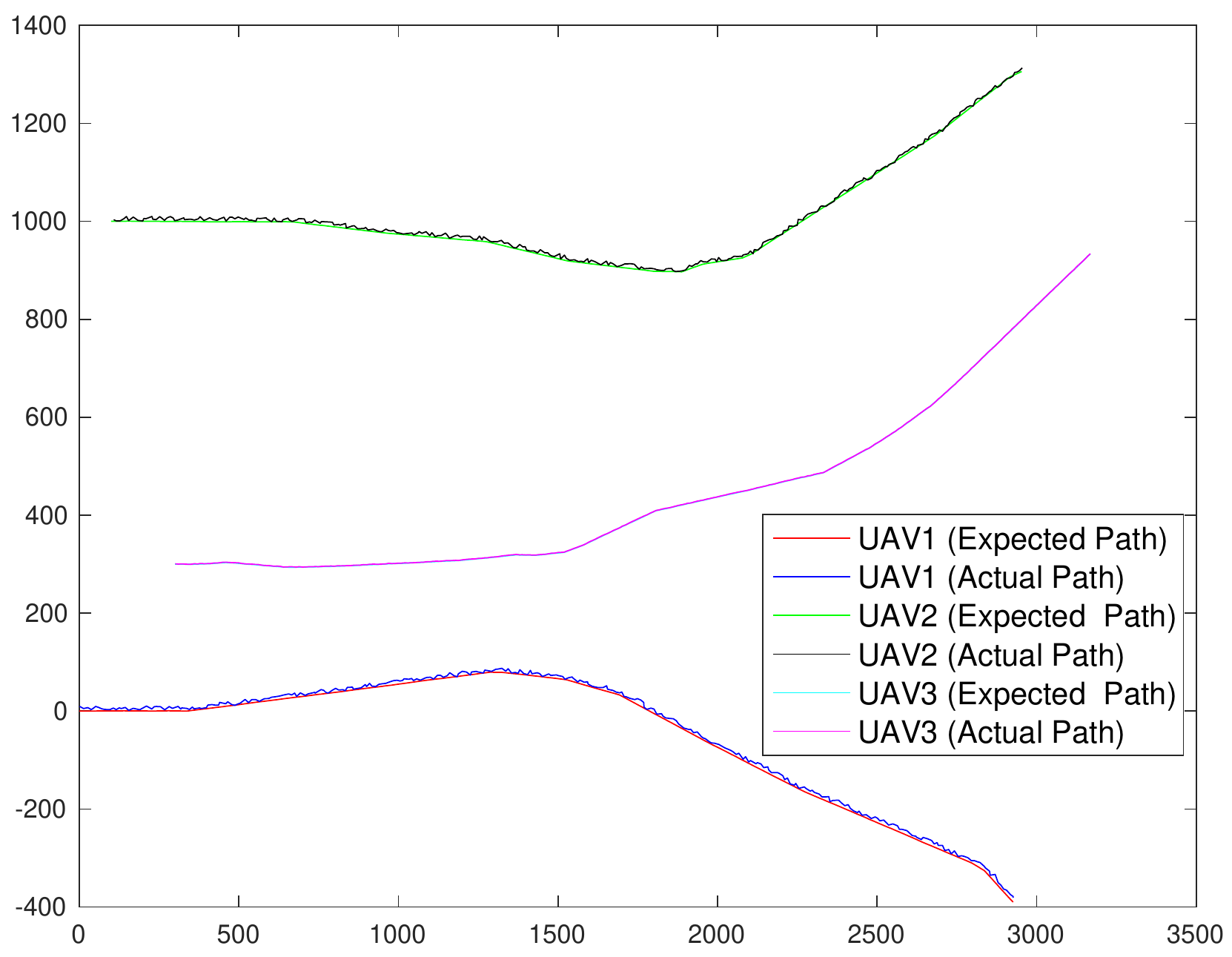}
         \caption{Flight trajectory of all UAVs compared to their predicted paths under a normal condition}
         \label{fig:fig2b}
     \end{subfigure}
        \caption{UAVs' behavior under a normal condition.}
        \label{fig:fig2}
\end{figure}

\begin{figure*}[!t]
	\centering
	\begin{subfigure}[b]{0.30\textwidth}
         \centering
         \includegraphics[width=\textwidth]{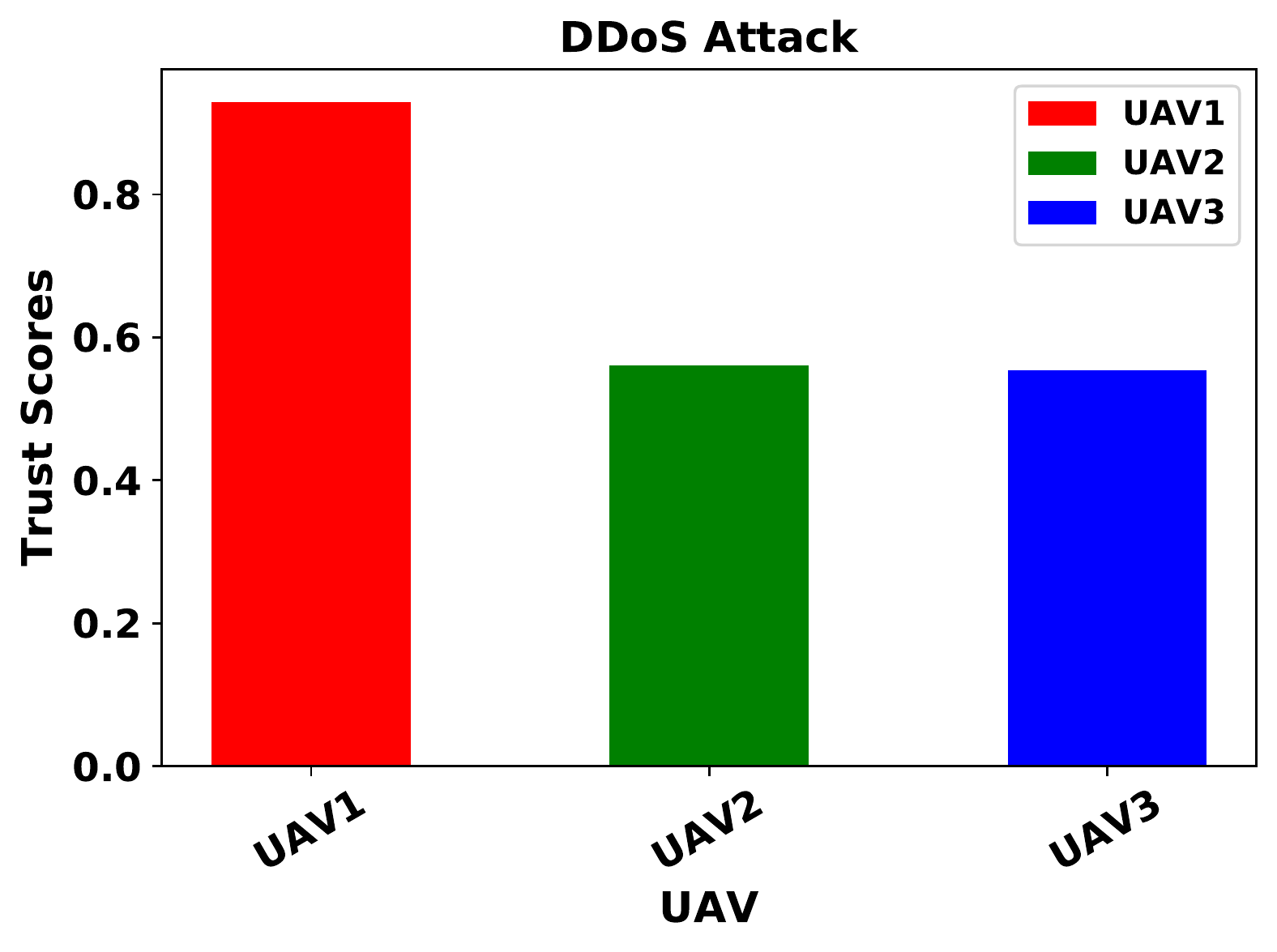}
         \caption{$UAV_1$ is under the DDoS attack}
         \label{subfig:fig3a}
     \end{subfigure}
     \hfill
     \begin{subfigure}[b]{0.30\textwidth}
         \centering
         \includegraphics[width=\textwidth]{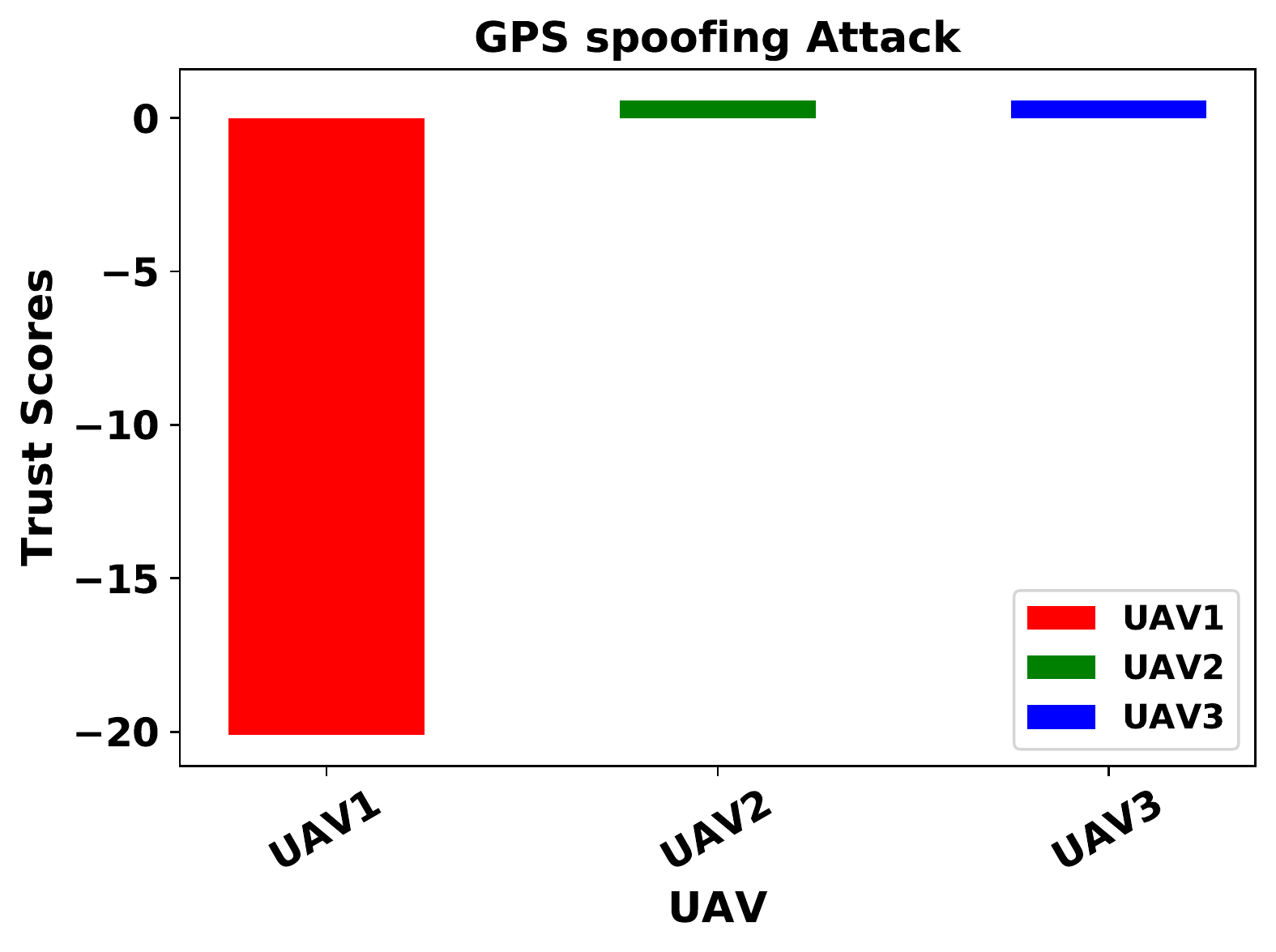}
         \caption{$UAV_1$ is under the GPS spoofing attack}
         \label{subfig:fig3b}
     \end{subfigure}
     \hfill
     \begin{subfigure}[b]{0.30\textwidth}
         \centering
         \includegraphics[width=\textwidth]{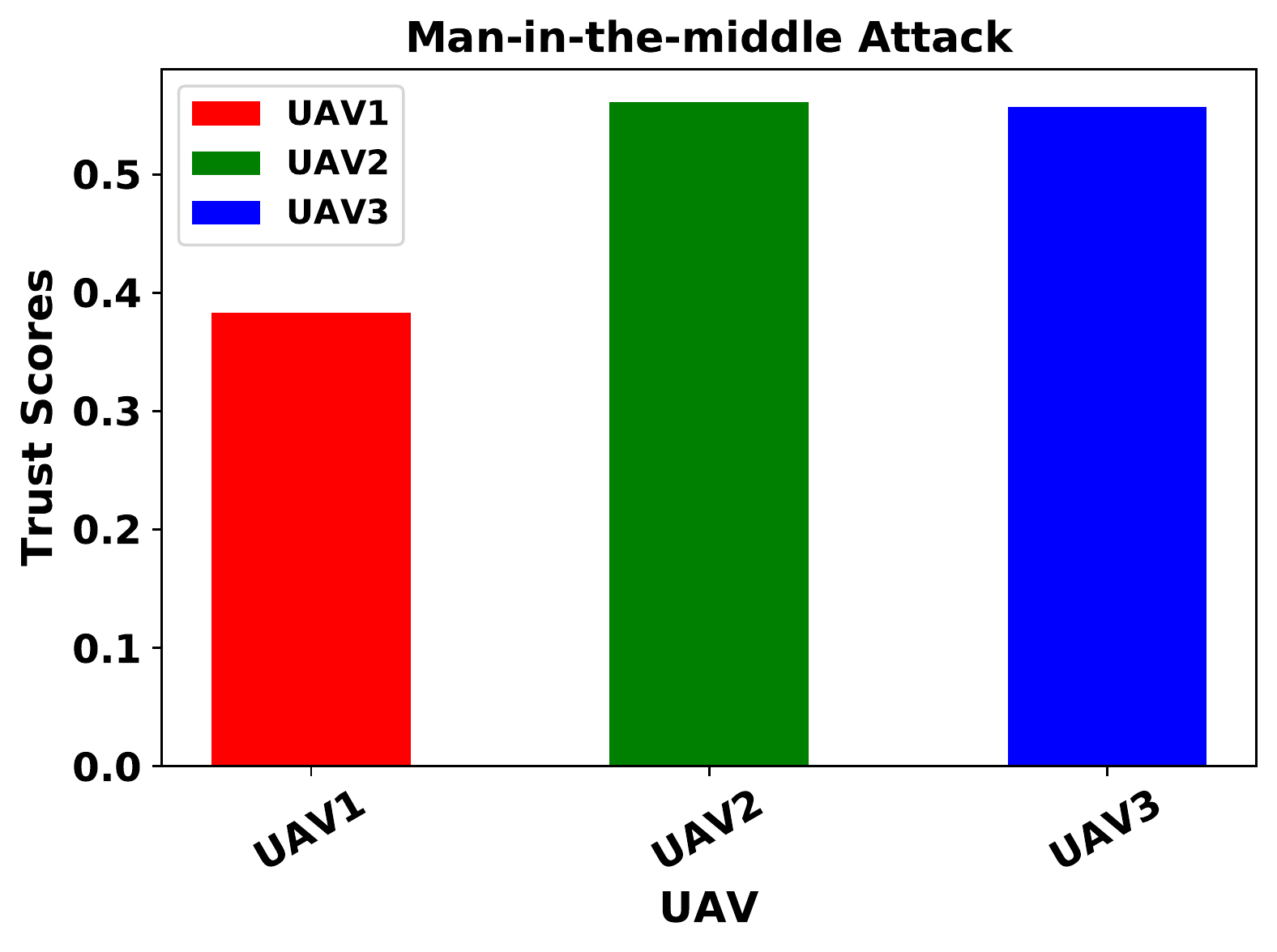}
         \caption{$UAV_1$ is under the man-in-the-middle attack}
         \label{subfig:fig3c}
     \end{subfigure}
     \\
     \begin{subfigure}[b]{0.30\textwidth}
         \centering
         \includegraphics[width=\textwidth]{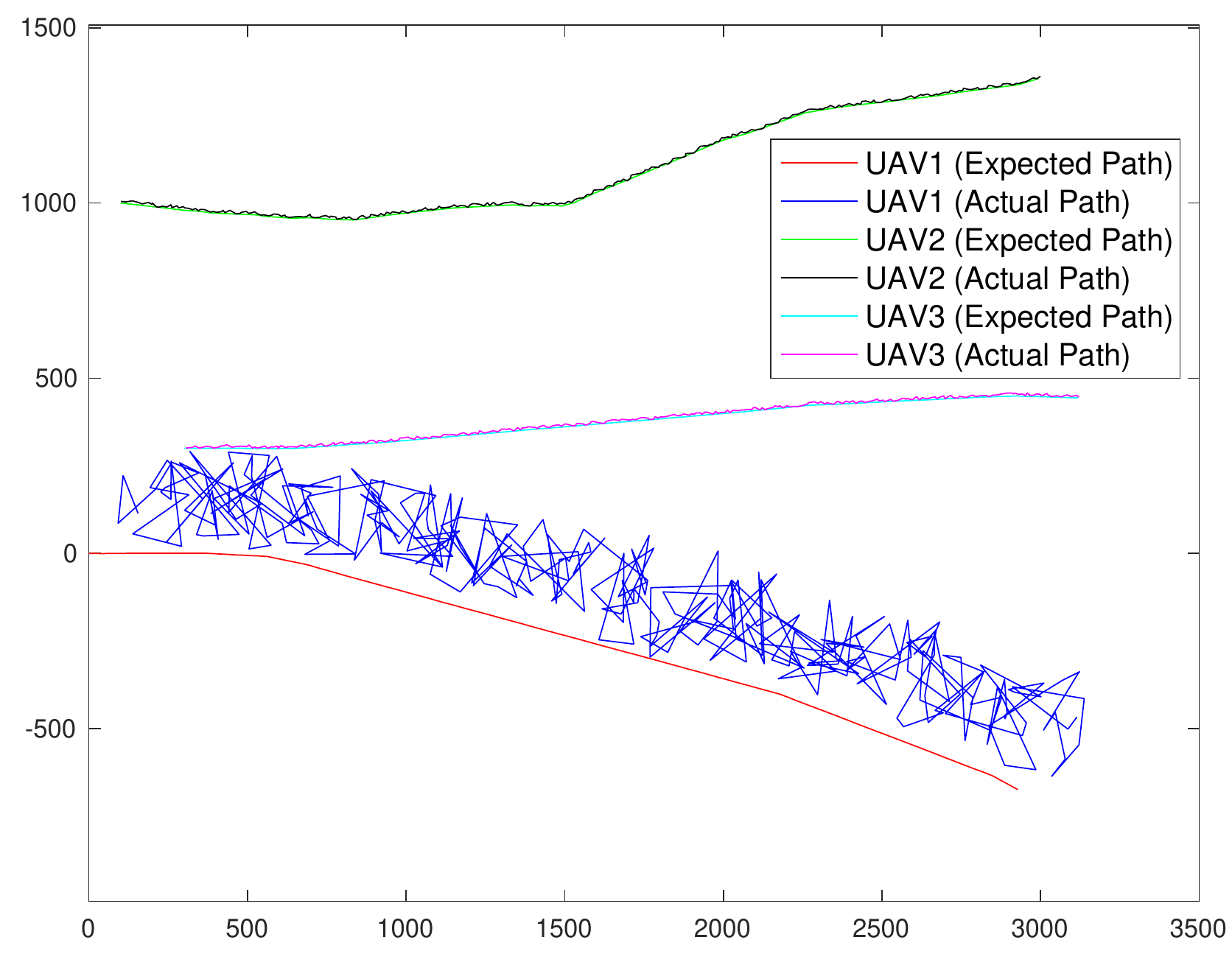}
         \caption{Flight pattern of $UAV_1$ under GPS spoofing attack}
         \label{fig:fig4b}
     \end{subfigure}
     \hfill
     \begin{subfigure}[b]{0.30\textwidth}
         \centering
         \includegraphics[width=\textwidth]{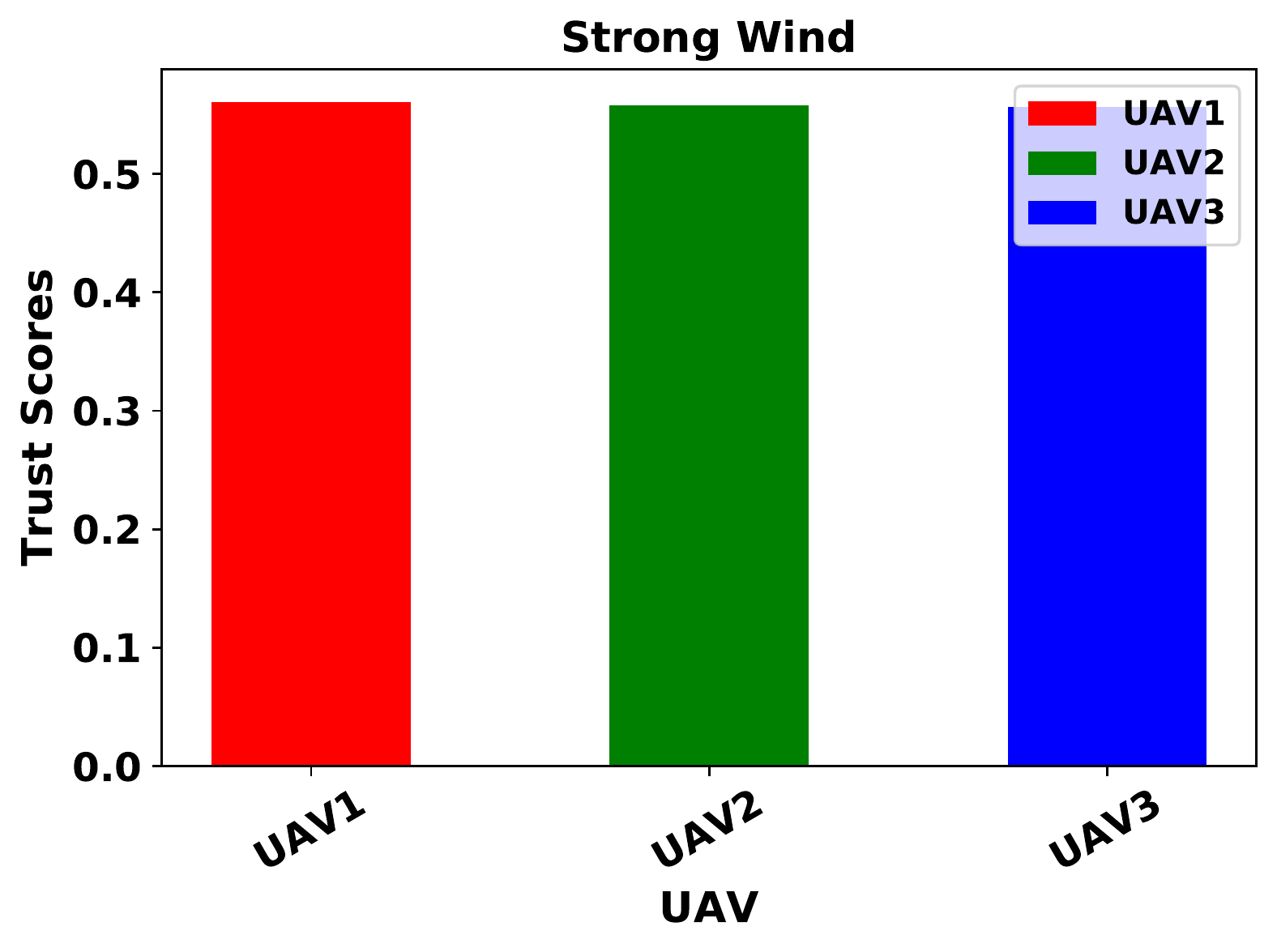}
         \caption{Trust score of all UAVs when experiencing a strong wind.}
         \label{fig:fig5}
     \end{subfigure}
     \hfill
     \begin{subfigure}[b]{0.30\textwidth}
         \centering
         \includegraphics[width=\columnwidth]{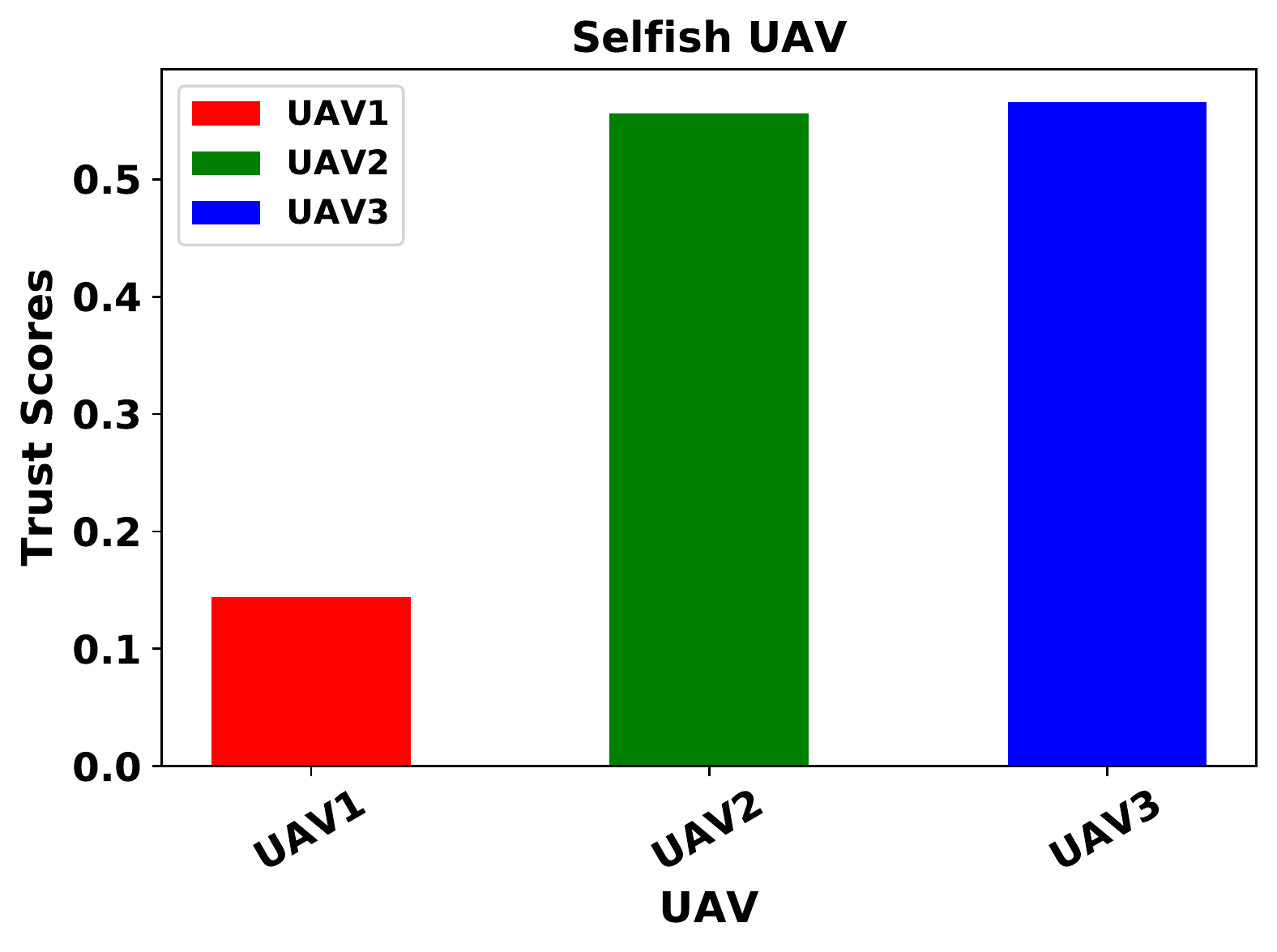}
         \caption{Trust scores of UAVs when $UAV_1$ acts selfishly or being hijacked.}
         \label{fig:fig4a}
     \end{subfigure}
     
    \caption{Trust scores and different scenarios for UAV$1$}
    \label{fig:fig3}
\end{figure*}
\begin{figure}[t]
     \centering
         \includegraphics[width=0.70\columnwidth]{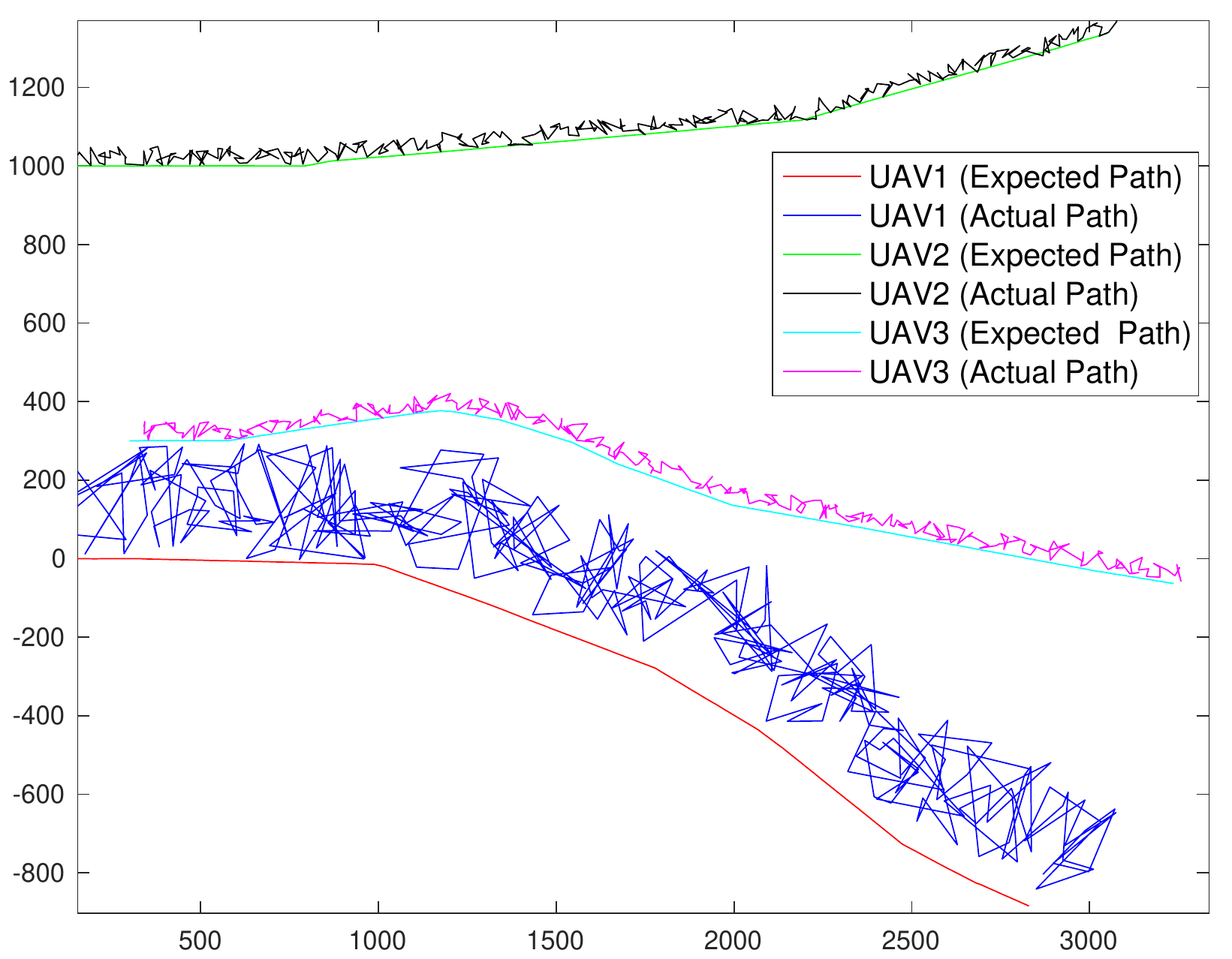}
         \caption{Wind and GPS Spoofing for flight patterns.}
         \label{fig:wind_attack}
\end{figure}

\begin{figure}[t]
     \centering
         \includegraphics[width=0.85\columnwidth]{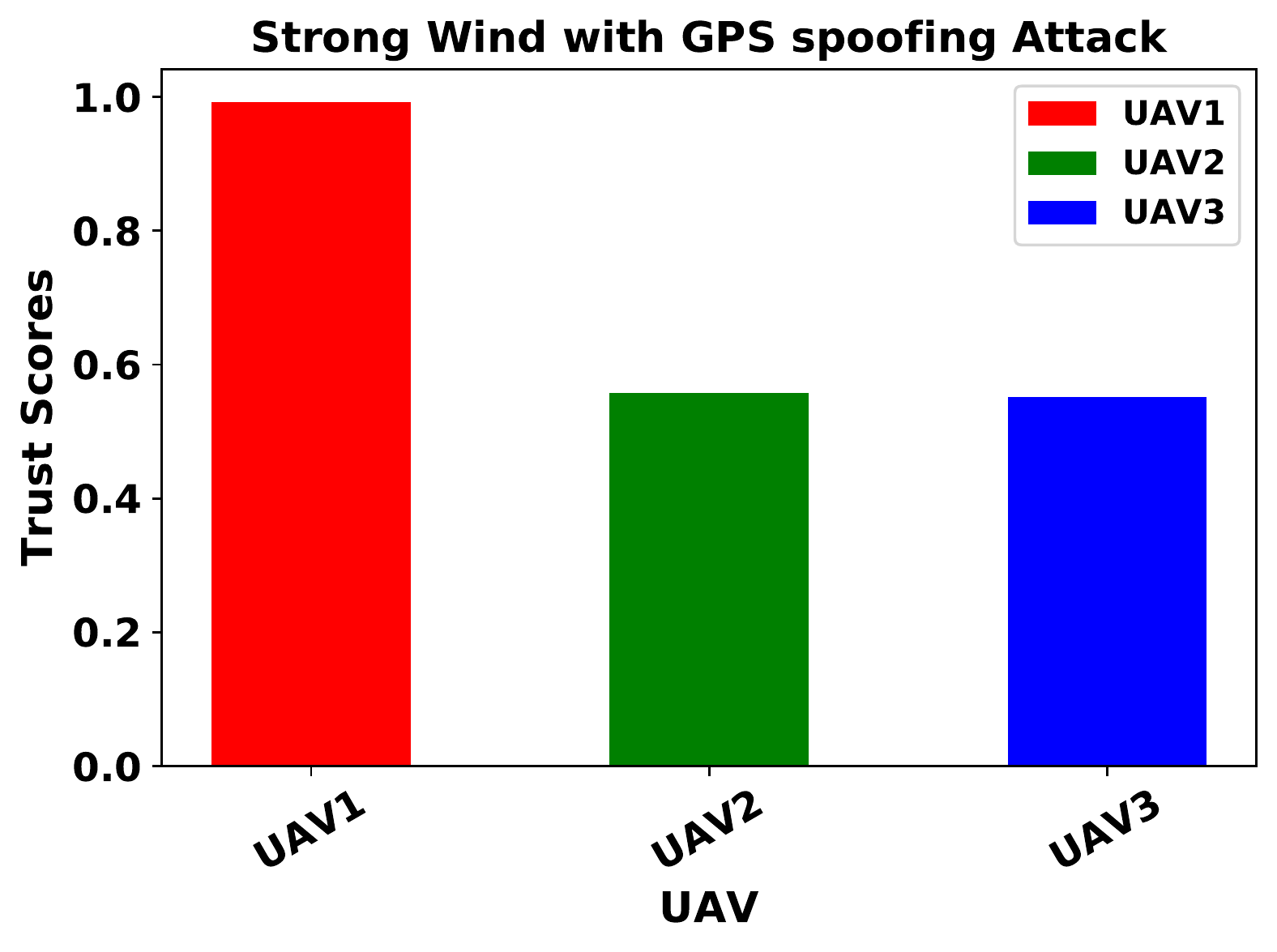}
         \caption{Trsut score of all UAVs in bad weather with GPS spoofing}
         \label{fig:wind_gps_score}
\end{figure}

\begin{algorithm}[hbtp]
\SetAlgoLined
 \For{all UAVs}
 {
    Calculate the task trust score based on (\ref{eq:trust_count})
    \\
    Calculate the energy trust score based on (\ref{eq:energy_trust})
    \\
    Calculate the deviation trust score based on (\ref{eq:trust_dev})
    \\
    And calculate the total trust score ($T_{i}$) based on {(\ref{eq:trust})}
 }
  \For{all UAVs}
 {
    \eIf{The range of the $T_i$s is different from other UAVs}
    {
          UAV$_i$ is not reliable
    }
    {
        No attack happened for UAV$_i$ 
    }
 }
 \caption{Trust model pseudo code for all UAVs}
 \label{algo:algorithm_trust}
\end{algorithm}

There are three different factors  of energy consumption, deviation the UAVs from their expected path, and the number of tasks that impact the trust score of each UAV.  Figure~\ref{fig:fig2} shows the performance of the UAVs in a normal condition where all of the trust scores are in the same range (Fig.~\ref{fig:fig2a}), and all the UAVs have a low-level of deviation from their predicted paths (Fig.~\ref{fig:fig2b}). 
Then, we consider  different attack scenarios to study whether our trust model can detect the attacks if at least one of the factors is impacted. In these scenarios, we assumed that $UAV_1$ is a malicious UAV that can be impacted by various attacks.

First, we consider a scenario where $UAV_1$ is under the DDoS attack. In this case, the energy consumption of $UAV_1$ is higher than the other UAVs operating in the same cluster. As shown in Fig.~\ref{subfig:fig3a}, the range of trust score for $UAV_1$ is higher than other  UAVs, since it has larger value of energy consumption.

In the second scenario, we study the case where $UAV_1$ is under the GPS spoofing attack. Under such attack, the UAV consumes more energy compared to the average  energy consumption rate, and it will not follow its expected path as shown in Fig.~\ref{fig:fig4b}. Noting (\ref{eq:trust}), the value of the deviation trust is a large number, hence the trust score is a negative number as shown in Fig.~\ref{subfig:fig3b}.

The third scenario is the man-in-the-middle attack, in which the number of failed tasks for the $UAV_1$ is higher than the number of successful tasks. In Fig.~\ref{subfig:fig3c}, the number of failed tasks is higher than the number of successful tasks for $UAV_1$. Based on (\ref{eq:trust_count}), the task trust is decreased, thus the trust score of $UAV_1$ is less than the other UAVs.

The fourth scenario refers to the case that $UAV_1$ has not delivered  its assigned tasks. This behavior can be an indicator that the UAV has been hijacked or show that the UAV is acting selfishly and does not follow the controller orders. This attack can translate to investing less energy in the assigned tasks compared to the average energy consumed by other UAVs. In Fig.~\ref{fig:fig4a}, the $UAV_1$ is selfish, hence the values of the task success rate and the energy consumption are small numbers, and consequently the trust score of this UAV is less than the other UAVs.



In the last scenario, we evaluate the performance of the proposed trust management mechanism in terms of understanding the difference between the UAVs' abnormal behavior when they are under an attack or when they are operating under harsh environmental conditions. First we test this scenario with the strong wind. In the strong wind scenario, all the UAVs consume more energy and show deviation from their predicted paths, therefore all of the trust scores are in the same range. We run this scenario 1000 times for different types of tasks and the average results show  that our model can correctly identify the strong wind with 70\% accuracy, as depicted in Fig.~\ref{fig:fig5}. Then, we test this scenario again by assuming that all the UAVs experience the strong wind while  UAV $i$ is also under the GPS spoofing attack. As we can see in Fig~\ref{fig:wind_attack}, all of the UAVs have deviation from their expected path, however $UAV_1$ has more deviation due to the GPS spoofing attack. Based on (\ref{eq:trust}) the trust score of the UAV under the attack is bigger than the rest of UAVs (shown in (\ref{fig:wind_gps_score})). Therefore, the proposed model can distinguish between the deviation of expected path due to the strong wind and the GPS spoofing attack.



The UAV node gets a signal from each DS. By subtracting the time the signal was transmitted from the time it was received, each DS can calculate how far it is from the UAV node. So given the travel time of the signals from the DS and the exact position of DSs, the position of the UAV node can be determined in three dimensions - east, north and altitude. As it it shown in equation \ref{eq:signal}, $t_i$ is the time that the UAV node received the signal, $s_i$ is the time the signal was transmitted by the $DS_i$, and $x_{i}, y_{i}, z_{i}$ shows the exact location of $DS_i$. $\beta$ is the receiver's clock bias from the much more accurate DS's clock.

\begin{align}\label{eq:signal}
& d_{(i,j)} = (t_{i} - \beta - s_i)*c
\end{align}

\begin{align}\label{eq:signal2}
& d_{(i,j)} = \sqrt{(x - x_{i})^2 + (y - y_{i})^2 + (z - z_{i})^2} ,
\end{align}

\section{Conclusion}
The problem of real-time detection of cyber-physical attacks in a network of drones is studied. We developed an online trust monitoring mechanism in which a central unit regularly observes the operations of the UAVs  in terms of their flight trajectory, their energy consumption as well as performing their assigned tasks and compare their trust factors to identify any potential abnormal behaviors. The proposed comparative approach enables the audit unit to differentiate between the abnormal behaviors due to the cyber-physical attacks or the potential unusual actions (e.g. turbulence or irregular energy consumption) due to facing harsh environmental conditions. Further, rather than relying on the self-report of the UAVs  to declare the number of their completed tasks or their future path, the proposed trust monitoring approach estimates the performance of the UAVs based on the observation of the audit unit while accounting for the potential uncertainty in such observations.



\bibliographystyle{IEEEtran}
\bibliography{main}

\end{document}